\documentclass[12pt]{article}

\usepackage{graphicx,setspace,float,rotating,longtable}
\usepackage{amsmath,verbatim,amssymb,epsfig}
\usepackage{color}
 \usepackage[breaklinks]{hyperref}
\usepackage{natbib}

\bibpunct{(}{)}{;}{a}{}{,}

\def\eqx"#1"{{\label{#1}}}
\def\eqn"#1"{{\ref{#1}}}


\begin{document}

\def\zengcomment#1{\vskip 2mm\boxit{\vskip 2mm{\color{red}\bf#1} {\color{blue}\bf --
D. Zeng\vskip 2mm}}\vskip 2mm}
\def\diaocomment#1{\vskip 2mm\boxit{\vskip 2mm{\color{blue}\bf#1} {\color{red}\bf --
G. Diao\vskip 2mm}}\vskip 2mm}

\def\dfrac#1#2{{\displaystyle{#1\over#2}}}
\def\boxit#1{\vbox{\hrule\hbox{\vrule\kern6pt
          \vbox{\kern6pt#1\kern6pt}\kern6pt\vrule}\hrule}}

\def\eqalign#1{\null\,\vcenter{\openup\jot\ialign
{\strut\hfil$\displaystyle{##}$&$\displaystyle{{}##}$\hfil\crcr#1\crcr}}\,}
\def\refhg{\hangindent=20pt\hangafter=1}

\baselineskip=24pt

\newcommand {\ttheta} {\mbox{\boldmath $\theta$}}
\def\refhg{\hangindent=20pt\hangafter=1}
\def\refmark{\par\vskip 2mm\noindent\refhg}
\def\bLambda{\mbox{\boldmath $\Lambda$}}
\def\bGamma{\mbox{\boldmath $\Gamma$}}
\def\bDelta{\mbox{\boldmath $\Delta$}}
\def\bdelta{\mbox{\boldmath $\delta$}}
\def\btheta{\mbox{\boldmath $\theta$}}
\def\bepsilon{\mbox{\boldmath $\epsilon$}}
\def\nbeta{\mbox{\boldmath $\eta$}}
\def\BTheta{\mbox{\boldmath $\Theta$}}
\def\bSigma{\mbox{\boldmath $\Sigma$}}
\def\bOmega{\mbox{\boldmath $\Omega$}}
\def\balpha{\mbox{\boldmath $\alpha$}}
\def\bLambda{\mbox{\boldmath $\Lambda$}}
\def\bpi{\mbox{\boldmath $\pi$}}
\def\bphi{\mbox{\boldmath $\phi$}}
\def\bPi{\mbox{\boldmath $\Pi$}}
\def\bxi{\mbox{\boldmath $\xi$}}
\def\ss#1{\scriptsize\uppercase{#1}}
\def\bJ{{\bf J}}

\newcommand{\bgamma}{\boldsymbol{\gamma}}
\newcommand{\bbeta}{\boldsymbol{\beta}}

\def\bmu{\mbox{\boldmath $\mu$}}
\def\bb{{\bf b}}
\def\ba{{\bf a}}
\def\bc{{\bf c}}
\def\be{{\bf e}}
\def\bg{{\bf g}}
\def\bh{{\bf h}}
\def\bl{{\bf l}}
\def\br{{\bf r}}
\def\bu{{\bf u}}
\def\bt{{\bf t}}
\def\bv{{\bf v}}
\def\bw{{\bf w}}
\def\bx{{\bf x}}
\def\by{{\bf y}}
\def\bz{{\bf z}}
\def\bA{{\bf A}}
\def\bB{{\bf B}}
\def\bC{{\bf C}}
\def\bE{{\bf E}}
\def\bG{{\bf G}}
\def\bH{{\bf H}}
\def\bI{{\bf I}}
\def\bJ{{\bf J}}
\def\bK{{\bf K}}
\def\bM{{\bf M}}
\def\bO{{\bf O}}
\def\bP{{\bf P}}
\def\bR{{\bf R}}
\def\bQ{{\bf Q}}
\def\bY{{\bf Y}}
\def\bU{{\bf U}}
\def\bX{{\bf X}}
\def\bY{{\bf Y}}
\def\bW{{\bf W}}
\def\bZ{{\bf Z}}
\def\bV{{\bf V}}
\def\bT{{\bf T}}
\def\bS{{\bf S}}
\def\bpsi{\mbox{\boldmath $\psi$}}

\renewcommand{\baselinestretch}{1.2}
\markright{
}
\markboth{\hfill{\footnotesize\rm Guoqing Diao
}\hfill}
{\hfill {\footnotesize\rm Semiparametric Hazards Rate Model} \hfill}
\renewcommand{\thefootnote}{}
$\ $\par
\vspace{0.8pc}
\centerline{\Large\bf Efficient Semiparametric Estimation of Short-term and}
\centerline{\Large\bf Long-term Hazard Ratios with Right-Censored Data}
\vspace{.4cm}
\begin{center}
{Guoqing Diao$^1$, Donglin Zeng$^2$ and Song Yang$^3$}\\
\vspace{.4cm}
{\it $^1$Department of Statistics, 
George Mason University\\
$^2$Department of Biostatistics, University of North Carolina at Chapel Hill\\
$^3$Office of Biostatistics Research, National Heart, Lung, and Blood Institute,
    National Institutes of Health}
\end{center}
\vspace{.55cm}

\begin{quotation}
\noindent {\it Abstract:}
The proportional hazards assumption in the commonly used
Cox model for censored failure time data is often violated in
scientific studies. Yang and Prentice (2005) proposed a novel
semiparametric two-sample model that includes the proportional
hazards model and the proportional odds model as sub-models, and
accommodates crossing survival curves. The model leaves the baseline
hazard unspecified and the two model parameters can be interpreted
as the short-term and long-term hazard ratios. Inference procedures
were developed based on a pseudo score approach. Although
extension to accommodate covariates was mentioned, no formal
procedures have been provided or proved. Furthermore, the pseudo
score approach may not be asymptotically efficient. We study the
extension of the short-term and long-term hazard ratio model of
Yang and Prentice (2005) to accommodate potentially time-dependent
covariates.
We develop efficient likelihood-based estimation and inference
procedures.
The nonparametric maximum likelihood
estimators are shown to be consistent, asymptotically normal, and asymptotically
efficient. Extensive simulation studies demonstrate that the
proposed methods perform well in practical settings. The proposed method
captured the phenomenon of crossing hazards in a cancer
clinical trial and identified a genetic marker with significant long-term
effect missed by using the proportional hazards model on age-at-onset of alcoholism in
a genetic study.

\par\vspace{9pt}
\par
\end{quotation}

\par\vspace{5mm}
\section{Introduction}
Much of the modern statistical methodology for survival analysis involves the seminar
work of \cite{Cox:72}.
The Cox proportional 
hazards model specifies that the hazard function of the event time
$T$ given a $p\times 1$ covariate vector $\bX$ takes the form
\begin{equation}
\lambda(t|\bX) = \lambda(t) e^{\bbeta^T\bX},
\label{Cox}
\end{equation}
where $\lambda(t)$ is an unspecified baseline hazard function and
$\bbeta$ is a $p\times 1$ vector of unknown regression parameters.
The assumption of constant relative risks over time in the Cox model, 
however, is often violated in many biomedical and genetic studies.
For instance, crossing
hazards may be  observed in clinical trials, in which the treatment has
certain adverse effect  initially but can be  beneficial in the long run.
In genetic studies, 
a certain gene may have a large impact on the hazard for children shortly 
after birth, but may have a relatively small impact later in life. 
In some other studies, 
genes related to susceptibility for a certain disease may affect older 
people more than younger people.
 
A motivating example is from the Collaborative 
Study on the Genetics of Alcoholism (COGA), a genetic family study with the aim 
of identifying and characterizing genetic factors that affect the susceptibility
to alcohol dependence and related phenotypes \citep{Hasin:03}.  
The investigators were particularly interested in assessing genetic effects
on the age at onset of ALDX1, the DSM-III-R+Feighner
classification status for alcohol dependence. 
Recent studies by \cite{Wang:06GE} and \cite{Diao:09GE} suggested that
SNP rs1972373 on chromosome 14 might be a disease susceptibility locus.
There are three possible genotypes, `1/1', `1/2', and `2/2', at SNP rs1972373. 
Kaplan-Meier estimates of survival curves for the
three genotype groups presented in Figure 1 appear to be overlapping with each other before
age of around 25, after that the curve for `1/1' begins to show more separation from the ones for the other two.  
In such situations, the proportional hazards 
model cannot distinguish short-term and long-term genetic effects. 
Another interesting example involves data from a randomized clinical trial on the treatment
of locally unresectable gastric cancer \citep{GTSG:82}. The aim of this trial 
was to compare chemotherapy with the combined chemotherapy and 
radiotherapy. As shown in \cite{Yang:05BKA} and \cite{ZengLin:07JRSSB}, 
the Kaplan-Meier survival curves for the two treatment groups 
cross at around 1000 days indicating crossing hazards.  The proportional 
hazards model
cannot capture crossing hazards and could yield very misleading results 
in such situations.

When the assumption of proportional hazards is questionable, an alternative
to the Cox model is the proportional odds model 
\citep{Bennett:83,Murphy:97JASA}, 
which assumes that the relative risk converges to one 
rather than remaining constant as time increases. The survival function 
of $T$ given covariates $\bX$ under the proportional odds model takes the 
form
\begin{equation}
S(t|\bX) = \frac{e^{-\bbeta^T\bX}}{G(t)+e^{-\bbeta^T\bX}},
\label{PO}
\end{equation}
where $G(\cdot)$ is a strictly increasing function with $G(0)=0$. 
Both the proportional
hazards and proportional odds models belong to the class of linear
transformation models which relate an unknown monotone transformation
of the failure time $T$ linearly to the covariates $\bX$ \citep
{Bickel:93,ZengLin:07JRSSB}.
The phenomenon of crossing hazards, however, cannot be directly captured 
by linear transformation models. 

\cite{Yang:05BKA} proposed a novel semiparametric two-sample hazard rate model that 
accommodates crossing survival curves. Their model leaves the baseline distribution
unspecified and the two model parameters have the appealing interpretations of the short-term 
and the long-term hazard ratios, respectively. The authors developed inference procedures based on
a pseudo score approach and showed that the estimators are consistent and asymptotically
normal. 
Although extension to accommodate covariates was mentioned, no formal procedures
have been provided or proved. 
In addition, the pseudo score approach may not be asymptotically efficient.

In this paper, we study the extension of the two-sample semiparametric hazard rate
model of \cite{Yang:05BKA} to accommodate covariates. Furthermore, the , covariates can be potentially time-dependent. 
We develop efficient likelihood-based estimation and inference procedures. The estimators
are shown to be consistent, asymptotically normal, and asymptotically efficient. 

The rest of the paper is organized as follows. In section 2, we introduce the 
semiparametric hazard rate model accommodating potentially time-dependent
covariates and formulate the 
nonparametric likelihood function. In Section 3, we describe the model
assumptions and derive the asymptotic results. 
Extensive simulations studies are presented in 
Section 4 to examine the finite sample properties of the proposed method. 
In Section 5, we illustrate
the new model through the applications to the gastric cancer trial and the COGA
study mentioned before. We conclude with a brief discussion in Section 6.
Proofs of the theoretical results are provided in the Appendix.

\section{Models and Inference}

Suppose that there is a random sample of $n$ independent subjects.
For the $i$th subject, let $T_i$ be the failure time, $C_i$ be the
censoring time, and $\bX_i$ be a $p\times 1$ vector of (time invariant)
covariates. The data consist of
$\{Y_i=\text{min}(T_i,C_i),\Delta_i=I(T_i\leq C_i), \bX_i,
i=1,...,n\}$, where $I(\cdot)$ is the indicator function. Let
$\tau$ be a constant denoting the end of the study.
We assume that $T_i$ and $C_i$ are independent given $\bX_i$.
We also assume that $P(C_i\geq \tau|{\bX}_i)
= P(C_i = \tau|{\bX}_i)>0$.

To incorporate short-term and long-term covariate effects, \cite{Yang:05BKA} 
discussed the following semiparametric hazard rate model
\begin{equation}
\lambda(t|\bX_i) = \frac{e^{(\bbeta+\bgamma)^T\bX_i}}
    {e^{\bbeta^T\bX_i}F(t)+
        e^{\bgamma^T\bX_i}S(t)}\lambda(t), 
\label{GH2}
\end{equation}
where $\lambda(t|\bX_I)$ is the hazard function of the event time $T_i$
given $\bX_i$, $\lambda(t)$ is the baseline hazard
function, $S(t)= \exp\{-\int_0^t \lambda(s) ds\}$ is the
baseline survival function, $F(t)=1-S(t)$ is the baseline cumulative
distribution function, and $\bbeta$ and $\bgamma$ are two vectors
of unknown regression parameters. The baseline cumulative hazard function
$\Lambda(t)\equiv \int_0^t \lambda(s)ds$ is left unspecified.
Under this model, the hazard ratios between two sets of covariate values are
allowed to be non-constant over time. Particularly, we can show that
\[
\lim_{t\rightarrow 0}\frac{\lambda(t|\bX_1)}{\lambda(t|\bX_2)}
	= e^{\bbeta^T(\bX_1-\bX_2)}, \ \
\lim_{t\rightarrow \tau_0}\frac{\lambda(t|\bX_1)}{\lambda(t|\bX_2)}
	= e^{\bgamma^T(\bX_1-\bX_2)},
\]
assuming the existence of the limits, where $\tau_0=\sup \{t:S(t)>0\}$.
Therefore, the parameters $e^{\bbeta}$ and $e^{\bgamma}$ can be
interpreted as the short-term and long-term hazard ratios, respectively.
Moreover, model \eqref{GH2} includes the proportional hazards
and proportional odds models as two sub-models, with
$\bbeta=\bgamma$ for the proportional hazards model \eqref{Cox},
and $\bgamma={\bf 0}$ for the proportional odds model \eqref{PO}.
 
We extend model \eqref{GH2} to allow time-dependent covariates. Let 
$\bX_i(\cdot)$ be a $p\times 1$ vector of
(possibly time-dependent) covariates. Also let $\overline{\bX}_i(t)$
denote the history of $\bX_i(\cdot)$ over $[0,t]$. We assume that the time dependent
covariates are external and that $\bX_i(\cdot)$
are bounded right-continuous functions with bounded right derivatives
in $[0,\tau]$ with probability one. We specify that the cumulative hazard function
conditional on $\overline{\bX}_i(t)$ takes the form
\begin{equation}
\Lambda(t|\overline{\bX}_i(t)) =
    \int_0^t \frac{e^{(\bbeta+\bgamma)^T\bX_i(s)}}
    {e^{\bbeta^T\bX_i(s)}F(s)+
        e^{\bgamma^T\bX_i(s)}S(s)}d\Lambda(s),
\label{GH1}
\end{equation}
where $\Lambda(t), S(t), F(t), \bbeta$, and $\bgamma$ have the same 
interpretation as those under model \eqref{GH2}.

Our goal is to make inference about parameters $\btheta\equiv(\bbeta,\bgamma)$ 
and the function $\Lambda(t)$.
Under the assumption of conditional independent censoring,
the likelihood for $(\btheta,\Lambda)$ takes the form
\[
\prod_{i=1}^n  \biggl[
    \frac{e^{(\bbeta+\bgamma)^T\bX_i(Y_i)}\Lambda'(Y_i)}
        {e^{\bbeta^T\bX_i(Y_i)}F(Y_i)+
        e^{\bgamma^T\bX_i(Y_i)}S(Y_i)}
    \biggr]^{\Delta_i} e^{ -\Lambda(Y_i|\overline{\bX}_i(Y_i))},
\]
where $\Lambda'(t)$ is the first derivative of $\Lambda(t)$.

In order to estimate the unknown parameters, we need to maximize the
observed-data likelihood. However, this maximum does not exist
because one can always choose $\Lambda'(Y_{i})=\infty$ for some $Y_{i}$
with $\Delta_{i}=1$.
Thus, we take a nonparametric maximum likelihood approach, in which
$\Lambda$ is allowed to be a right-continuous function. Specifically, we
replace $\Lambda'(Y_i)$ with $\Lambda\{Y_i\}$, the jump size of $\Lambda(\cdot)$
at $Y_{i}$. Therefore, 
we obtain the following nonparametric likelihood function
\begin{equation}
L_n(\btheta,\Lambda)=\prod_{i=1}^n  \biggl[
    \frac{e^{(\bbeta+\bgamma)^T\bX_i(Y_i)}\Lambda\{Y_i\}}
        {e^{\bbeta^T\bX_i(Y_i)}F(Y_i)+
        e^{\bgamma^T\bX_i(Y_i)}S(Y_i)}
    \biggr]^{\Delta_i} e^{ -\Lambda(Y_i|\overline{\bX}_i(Y_i))}.
\label{NonparaLikelihood}
\end{equation}

We maximize the nonparametric log-likelihood function $l_n(\bphi)\equiv \log L_n(\bphi)$.
The resultant nonparametric maximum likelihood estimators (NPMLEs) are denoted by $(\widehat{\btheta}_n,\widehat{\Lambda}_n)$.
It is easy to show that $\widehat{\Lambda}_n$ must be a step function
with positive jumps only at the $Y_i$s for which $\Delta_i=1$. We order the
distinct observed failure time as $(Y_{(1)},...,Y_{(m)})$, where $m$ is the total 
number of distinct observed failure times. Therefore, the above maximization
should be performed over the parameters $\btheta$ and these positive jumps. The cumulative
hazard function
$\Lambda(t|\overline{\bX}_i(t))$ in \eqref{NonparaLikelihood} takes the form
\[
\sum_{k:Y_{(k)}\leq t}\frac{e^{(\bbeta+\bgamma)^T\bX_i(Y_{(k)})}}
    {e^{\bbeta^T\bX_i(Y_{(k)})}F(Y_{(k)})+
       e^{\bgamma^T\bX_i(Y_{(k)})}S(Y_{(k)})}\Lambda\{Y_{(k)}\}.
\]

To compute the NPMLEs, 
we use the quasi-Newton algorithm described in Chapter 10 of \cite{Press:92}.
Specifically, we use the Broyden-Fletcher-Goldfarb-Shanno (BFGS) method,
which is one of the most efficient method for solving nonlinear optimization problems, and was proposed by \cite{Broyden:1970}, 
\cite{Fletcher:1970}, \cite{Goldfarb:1970}, and \cite{Shanno:1970}
individually. 
The BFGS method and its variants have been implemented in standard software such as
SAS, R, and Matlab and have been successfully used in literature. 
To ensure the stability of the quasi-Newton algorithm, we suggest to center covariates at their means.
When we constrain the
regression parameters such that $\bbeta=\bgamma$, the quasi-Newton algorithm yields 
the exactly the same parameter estimates as those from the procedure $phreg$ in SAS software and R routine
$coxph$ under the proportional hazards model; when we constrain $\bgamma={\bf 0}$,  the NPMLEs obtained from the quasi-Newton
algorithm are the same as those from R routine $nltm$ under the proportional odds model. 
These results provide an empirical validation of the quasi-Newton algorithm. 

In the next section, we will establish consistency and asymptotic normality of the NPMLEs.
We will show that the asymptotic covariance matrix for $\widehat{\btheta}_n$ attains the 
semiparametric efficiency bound and can be consistently estimated using the inverse of the observed Fisher information matrix for all parameters including $\btheta$ and the jump sizes of
$\widehat{\Lambda}_n$. 
Alternatively, following the argument of \cite{Murphy:00},
we can estimate the covariance matrix of $\widehat{\btheta}_n$ by using the 
profile likelihood function for $\btheta$, which is defined as the 
maximum likelihood of $L_n(\btheta,\Lambda)$ for any fixed $\btheta$. 
Our simulation studies indicated that both approaches work very well in
practical situations.

The formulation of the semiparametric hazard rate model provides an appealing 
diagnostic tool for testing the proportional hazards and proportional odds models 
since the latter two models are embedded in the former. Specifically, we can check 
the proportional hazards and proportional odds assumptions by testing $H_0:\bbeta=\bgamma$ 
and $H_0:\bgamma={\bf 0}$, respectively. This can be done by the Wald, score or
likelihood ratio statistics.

\section{Asymptotic Properties}

Let $\btheta_0=(\bbeta_0,\bgamma_0)$ and $\Lambda_0$ denote the true values of
$\btheta$ and $\Lambda$. We impose the following
regularity conditions:
\begin{enumerate}
\item[(C1)]
With probability one, the covariates $\overline{\bX}_i$ possess bounded total variation
in $[0,\tau]$ and the support of $\overline{\bX}_i$ contains 0. In addition, if there exists a function $c_0(t)$
and a constant vector $\bc_1$ such that
\[
\bc_1^T\bX_i(t)=c_0(t), \forall t\in[0,\tau]
\]
with probability one, then $c_0(t)=0$ and $\bc_1={\bf 0}$.

\item[(C2)] Conditional on $\overline{\bX}_i$, the censoring
time $C_i$ is independent of the failure time $T_i$.

\item[(C3)] There exists some positive constant number $\delta_0$
such that $P(C_i\geq \tau|\overline{\bX}_i) = P(C_i=\tau|\overline{\bX}_i)\geq
\delta_0$ almost surely, where $\tau$ is a constant denoting
the end of the study.

\item[(C4)] The true parameter value of $\btheta$, $\btheta_0$, belongs to a known
compact set $\mathcal{B}_0$ in $R^{2p}$.

\item[(C5)] The true baseline cumulative distribution function
$\Lambda_0$ belongs to the following class
\[
\begin{split}
\mathcal{A}_0 = & \{\Lambda: \Lambda \text{ is a strictly increasing function in } [0,\tau]
\text{ and is continuously differentiable}\\
	&\text{ with }\Lambda(0)=0, \Lambda'(0)>0\text{ and }
\Lambda(\tau)<\infty\}.\\
\end{split}
\]
\end{enumerate}

All the above assumptions are standard in the semiparametric 
analysis of failure time data. Under these assumptions, we 
first show that the NPMLEs $(\widehat{\btheta}_n,
\widehat{\Lambda}_n)$ exist. It suffices to show that the jump size of $\widehat{\Lambda}_n$ at
$Y_i$ for which $\Delta_i=1$ is finite. By the compactness of 
$\btheta$, $F$, $S$, and 
$\overline{\bX}_i, i=1,...,n$, we have
\[
L_n(\btheta,\Lambda) \leq
\prod_{i=1}^n c_1 \Lambda\{Y_i\}^{\Delta_i} e^{-c_2 \Lambda\{Y_i\}}
\]
for some positive constants $c_1$ and $c_2$. 
Thus, if for some $i$ such that $\Delta_i=1$ and 
$\Lambda\{Y_i\}\rightarrow\infty$, $L_n(\btheta,\Lambda)\rightarrow 0$. We conclude that the jump
sizes of $\widehat{\Lambda}_n$ must be finite. On the other hand, $\btheta$ belongs to a
compact set $\mathcal{B}_0$. It follows that the NPMLEs exist.

We next establish identifiability of the model parameters $(\btheta,\Lambda)$.

{\bf Lemma 1.} Under conditions (C1) - (C5), the parameters $\btheta$
and $\Lambda$ are identifiable.

The proof of Lemma 1 is given in Appendix A.1. 
Using Lemma 1, we are able to obtain the following consistency results.

{\bf Theorem 1.} Under conditions (C1)-(C5), $||\widehat{\btheta}_n-\btheta_0||\rightarrow 0$ and
$\underset{t\in[0,\tau]}{\sup}|\widehat{\Lambda}_n(t)-\Lambda_0(t)|
\rightarrow 0$ almost surely, where $||\cdot||$ is the Euclidean norm.

{\it Remark 1.} Theorem 1 states the consistency of the NPMLEs. The basic
idea to prove Theorem 1 is as follows. As in the proof of the existence of the NPMLEs, 
we will show that $\widehat{\Lambda}_n(\tau)$ is not allowed to diverge. Once the boundedness
of $\widehat{\Lambda}_n(\tau)$ is established, a subsequence of
$\widehat{\Lambda}_n$ can be found to converge pointwise to a
bounded monotone function $\Lambda^*$ in $[0,\tau]$ and the same subsequence
of $\widehat{\btheta}_n$ converges to some $\btheta^*$.
We construct a step function $\overline{\Lambda}_n$ with jumps at the 
observed failure times converging to $\Lambda_0$.
Then, because $L_n(\widehat{\btheta}_n,\widehat{\Lambda}_n)\geq L_n(\btheta_0,\overline{\Lambda}_n)$,
by taking the limit, we will prove that the Kullback-Leibler information
between the true density and the density indexed by $(\btheta^*,\Lambda^*)$
is non-positive. Therefore, the true density must be equal to the density
indexed by $(\btheta^*,\Lambda^*)$. The consistency
will then follow from the identifiability result. The detail of the proof is 
given in Appendix A.2.

Our last theorem establishes the asymptotic properties of the NPMLEs.

{\bf Theorem 2.} Under conditions (C1)-(C5), the random element
$\sqrt{n}(\widehat{\btheta}-\btheta_{0},
\widehat{\Lambda}_n-\Lambda_0)$ converges weakly to a zero
mean Gaussian process in the metric space $l^\infty(\mathcal{H})$,
where
\[
\mathcal{H}=\{(\bh_1,\bh_2,h_2):\bh_1\in R^{p},\bh_2\in R^p, h_3 \text{ is a function
on }[0,\tau]; ||\bh_1|| \leq 1, ||\bh_2||\leq 1, |h_3|_V\leq 1\}
\]
and $|h_3|_V$ denotes the total variation of $h_3$ in $[0,\tau]$.
Furthermore, $\widehat{\btheta}_n$ is asymptotically efficient.

{\it Remark 2.} In the statement of Theorem 2, asymptotically efficient 
estimators mean that the asymptotic covariances attain the semiparametric 
efficiency bounds as defined in \cite{Bickel:93}. Once the consistency
of the NPMLEs is established, the asymptotic distribution of the NPMLEs
stated in Theorem 2 can be derived by verifying the four conditions in
Theorem 3.3.1 of \cite{wellner:96}. 
The proof of Theorem 2 is given in Appendix A.3. 

{\it Remark 3.} Theorem 2 implies that for any $(\bh_1, \bh_2, h_3)\in \mathcal{H}$, 
$\sqrt{n}(\widehat{\bbeta}_n-\bbeta_0)^T\bh_1+
\sqrt{n}(\widehat{\bgamma}_n-\bgamma_0)^T\bh_2+\sqrt{n}\int_0^\tau
h_3(t)d(\widehat{\Lambda}_n-\Lambda_0)$ is asymptotically
normal with mean zero and variance $\text{Var}(\Psi[\bh_1,\bh_2,h_3])$, and
this normal approximation is uniform in $(\bh_1,\bh_2,h_3)$,
where $\Psi\in \infty(\mathcal{H})$ is the random element in the limiting distribution. 
Therefore, to estimate the variance of $(\widehat{\bbeta}_n,\widehat{\bgamma}_n,
\widehat{\Lambda}_n)$, we view \eqref{NonparaLikelihood} as a parametric likelihood
with $\bbeta,\bgamma$, and the jump sizes of $\Lambda$ at the observed failure times
as parameters. We can then estimate the asymptotic variance matrix of the unknown 
parameters by inverting the observed information matrix according to the parametric 
likelihood theory.

\section{Simulation Studies}
We conducted extensive simulation studies to evaluate the finite sample
performance of the proposed methodology using 1000 replicates. 
We generated failure times from the following model
\[
\Lambda(t|X_i)  = \int_0^t\frac{e^{(\beta+\gamma)X_i}}
    {e^{\beta X_i}F(s)+
        e^{\gamma X_i}S(s)}d\Lambda(s), 
\]
where $X_i$ is a uniform$(-1,1)$ variable.  
The baseline cumulative hazard function
is set to be $\Lambda(t)=t$. We consider four scenarios for the values of
regression parameters: (a) $(\beta,\gamma)=(-0.5,0.5)$; (b)
$(\beta,\gamma)=(-0.5,0)$; (c) $(\beta,\gamma)=(0,0.5)$;
and (d) $(\beta,\gamma)=(0.5,0.5)$. Under scenario (a), the short-term
and long-term hazard ratios are on opposite directions; under scenario (b),
the long-term hazard ratio is 1 corresponding to a true proportional odds model;
under scenario (c), the short-term hazard ratio is 1; and under scenario (d),
the short-term and long-term hazard ratios are equal corresponding to 
a true proportional hazards model. The censoring time is set to be the minimum of 
2 and a uniform$(0,4)$ variable, producing approximately 29\% censoring under all
four scenarios. We used the quasi-Newton algorithm \citep{Press:92} to
calculate the NPMLEs. There is little difference between the standard error estimates
through the Fisher information matrix and those from the profile likelihood approach.
We present the standard error estimates based on the observed 
Fisher information matrix throughout the simulation studies and real data applications.

Table 1 summarizes the results for $\beta, \gamma$, and $\Lambda(t)$ with $n=100$ 
and $n=200$. 
For the nonparametric estimation of 
$\Lambda(t)$, we evaluated its estimates at $t=0.5$ and $t=1.0$. 
For comparison, we also fit the proportional hazards and proportional 
odds models, for which the regression parameters were denoted as $\beta_{PH}$
and $\beta_{PO}$, respectively. The results in Table 1 indicate that the 
proposed method performs well for small sample sizes. In particular, 
the proposed estimators appear to be unbiased. The standard error estimator
reflects accurately the true variation, and the confidence intervals
have proper coverage probabilities. 
When the proportional hazards assumption is violated, the Cox model leads to 
biased estimates. Particularly, the results based on the Cox model can be
very misleading when the short-term and long-term covariate effects are 
in opposite directions. Similar results were observed for the proportional
odds model when the model assumption is not true. 
When the Cox model or the proportional odds model holds, as expected, 
the proposed NPMLEs are less efficient than those obtained under the 
true sub-model. 

Our next set of studies evaluated the proposed inference procedures
for the testing of covariate effects and the assumptions of proportional 
hazards and proportional odds. 
Specifically, we considered Wald tests for 
the following null hypotheses: (H1) $H_0: \beta=0$;
(H2) $H_0: \gamma=0$; 
(H3) $H_0: \beta=\gamma=0$; and (H4) $H_0: \beta=\gamma$.
Note that testing the long-term hazard ratio is equivalent to 
testing the proportional odds model. For comparison, we also
considered the testing of covariate effects under the proportional
hazards model: (H5) $H_0:\beta_{PH}=0$.
We used the same simulation setting as above with $n=200$. Table 2 
presents the sizes/powers of the Wald tests at the nominal
levels of 0.05. 
In all cases, the proposed 
tests have accurate control of type I error rates and reasonable powers under the
alternative. The proposed tests of short-term, long-term and overall 
covariate effects tend to be more powerful than the Cox model when the
proportional hazards assumption is violated.
When our interest is to test
the short-term or long-term hazard ratio only, the Cox model 
tends to yield inflated type I error rates under model mis-specifications.
 
 We carried out additional simulation studies to compare the efficiency of 
 the proposed NPMLEs relative to 
 the pseudo-maximum likelihood estimators for two-sample data as 
 implemented by \cite{Yang:05BKA}. We considered the same simulation
 settings as above except that $X_i$ is a binary variable taking values -0.5 and 0.5 
 with equal probabilities. Table 3 presents the empirical mean squared errors 
 for estimating $\beta$ and $\gamma$ based on 1,000 repetitions. 
 As expected, under almost all situations
 the proposed estimators are more efficient than the pseudo-maximum likelihood
 estimators.

\section{Real Data Examples}

\subsection{COGA study}
In the COGA study mentioned previously, 643 individuals 
were affected with alcoholism and 971 individuals were disease-free
at the time of interview. After excluding individuals with missing 
genotype at the target gene locus or phenotype data, the final data set for 
our analysis consisted of 1,371 individuals, including 626 affected individuals 
and 745 unaffected individuals.

Preliminary analysis revealed that gender was a risk 
factor for alcoholism; males were at a higher risk than females. 
Of the 626 affected individuals, 424 were males, as opposed 
to 229 males in the unaffected individuals. 
Previous linkage analysis showed a linked region 
on chromosome 14 \citep{Palmer:99}. Two recent studies on the genetic 
association analysis of ordinal traits \citep{Wang:06GE, Diao:09GE}
suggested that SNP rs1972373 on chromosome  
14 might be a disease susceptibility locus. 
Based on the Kaplan-Meier estimates of survival curves for the three genotype groups
at SNP rs1972373 presented in Figure 2, allele `2' appeared to have little short-term impact
but strong long-term impact on the risk of alcoholism. 

In our analysis, we fit the proposed model \eqref{GH1} and included gender and
genotype score at SNP rs1972373 as covariates. 
The gender of an individual was coded as 1 for male and 0 for female, and the 
genotype score was coded as the numbers of allele type `2'.
Both covariates were then centered at
their means. The tests of
the proportional hazards assumption for gender and genotype score at
SNP rs1972373 were significant with p-values of 0.016 and 0.027.
Gender appeared to have significant short-term
and long-term effects on the age-at-onset of alcoholism. 
The short-term and long-term log-hazard ratios of male versus female
are estimated at 0.866 and 1.9932 with standard error estimates
of 0.147 and 0.367, both leading to $p$-values less than 0.0001. 
As expected, SNP rs1972373 appeared to have no short-term effect
but significant long-term effect on the age-at-onset of alcoholism.
The short-term log-hazard ratio of allele type `2' versus allele type 
`1' is estimated at -0.06 with a p-value of 0.479 whereas the 
long-term log-hazard ratio is estimated at 0.683 with a p-value 
of 0.015. One copy of allele type `2' in the genotype at SNP
rs1972373 is expected to increase the long-term hazard of alcoholism by
98\% with a 95\% confidence interval of $(14\%,243\%)$. Figure 1 plots
the separate Kaplan-Meier and the model-fitted survival curves for
each genotype group. The model-fitted survival function is calculated
as the empirical average of the predicted survival functions. That
the predicted survival functions agree well with the nonparametric
Kaplan-Meier estimates of the survival curves indicates a good fit
of the model. In contrast, the Cox model failed to detect 
the long-term effect of SNP rs1972373.
The log-hazard ratio estimated from the Cox model is 0.083 with 
a standard error estimate of 0.058, corresponding to a p-value of 
0.153. 

\subsection{Gastrointestinal tumor study}
As mentioned in the Introduction section, the gastrointestinal tumor study 
compared chemotherapy with the combined chemotherapy and radiotherapy
on the treatment of locally unresectable gastric cancer. 
There were 45 patients randomly assigned to each treatment arm. 
Two observations were censored in the 
chemotherapy group and six were censored in the combined therapy group. 
Under the two-sample proportional hazards model, the log-hazard ratio of
chemotherapy versus the combined therapy is estimated at 0.106 with a standard
error estimate of 0.223, yielding a $p$-value of 0.635. The use of proportional
hazards model failed to capture the phenomenon of crossing survival curves 
shown in Figure 1 and the results were meaningless in this situation. 

We fit the proposed model \eqref{GH1} by letting $X_i=0.5$ for the combined therapy group
and $X_i=-0.5$ for the chemotherapy group. 
The test of the proportional hazards 
assumption is highly significant with a p-value of $6.0\times 10^{-4}$.
The new method successfully
captured the phenomenon of crossing hazards.
The short-term log-hazard ratio 
$\beta$ and long-term log-hazard ratio $\gamma$ are on opposite
directions and estimated at 1.76 and
-1.59 with standard error estimates of 0.582 and 0.509, leading to $p$-values
of 0.0025 and 0.0018, respectively. The 95\% confidence intervals are
$(0.62,2.90)$ for $\beta$ and $(-2.59,-0.59)$ for $\gamma$. The estimated
short-term and long-term hazard ratios are 5.81 and 0.20 with 95\% confidence
intervals $(1.86,18.17)$ and $(0.075,0.553)$. As evident in Figure 2, the
model fitted survival curves agree well with the nonparametric Kaplan-Meier
survival estimates very well indicating a good model fit. Our results are also 
consistent with the results from the two-sample model of \cite{Yang:05BKA} using the
pseudo maximum likelihood approach.

\section{Discussion}
We have extended the two-sample semiparametric hazard rate model of \cite{Yang:05BKA} to incorporate
short-term and long-term effects of potentially time-dependent
covariates. We have studied the nonparametric maximum likelihood
estimation for the proposed model \eqref{GH1} and established the asymptotic
properties for the NPMLEs. 
Unlike the existing varying-coefficient 
Cox model, the estimation and inference procedures are likelihood-based
and statistically efficient.
Numerical studies and the applications to the Gastrointestinal 
tumor study and the COGA study demonstrate that the proposed
inference procedures perform well in practical situations.

We have implemented the new method in C language using the
quasi-Newton algorithm described in \cite{Press:92}. The convergence of the 
quasi-Newton algorithm is very fast and it takes less 
than 0.2 second to analyze one data set with 400 subjects 
on a Dell PowerEdge 2900 server. The efficiency 
of our computer program makes it feasible to apply our method 
to gene expression data and genome-wide association studies. 
Our user-friendly computer program is 
freely available on the 
website: http://mason.gmu.edu/$\sim$gdiao/software/.

For the purpose of illustration, we assume that observations in the COGA study
are independent. Although the failure times within the same family tend to be
correlated, the NPMLEs $\widehat{\btheta}_n$ can be shown to be consistent for
$\btheta$ and asymptotically normally distributed provided that the marginal model
is corrected specified. However, the naive covariance matrix estimator 
for $\widehat{\btheta}_n$ using the inverse of the observed Fisher information matrix,
is no longer valid in the presence of within-family dependence. To account for
within-family correlations, one option is to fit marginal models 
and then use the robust sandwich estimators of covariance matrix. For the COGA data,
the naive and robust covariance estimates were very close suggesting weak 
within-family correlations. Currently we are investigating the extensions of the semiparametric 
hazard rate model \eqref{GH1} to correlated failure time data by using random effects.

To assess the adequacy of the semiparametric hazard rate model \eqref{GH1}, we can develop a goodness-of-fit procedure based on martingale residuals.
The martingale under model \eqref{GH1} can be written as
\[
M_i(t) = N_i(t) - \int_0^t Y_i(s) \frac{e^{(\bbeta+\bgamma)^T\bX_i(s)}}
        {e^{\bbeta^T\bX_i(s)}F(s)+e^{\bgamma^T\bX_i(s)}S(s)}d\Lambda(s),
\]
where $N_i(t)$ and $Y_i(t)$ are the usual counting process
and at risk process. The score process for $\btheta$ seen as a function of
time can be expressed as functions of martingale residuals,
\[
\bU(t;\btheta,\Lambda)
= \sum_{i=1}^n \int_0^t \bZ_i(s) dM_i(s),
\]
where
\[
\bZ_i(s) = \begin{bmatrix}
        \pi_i(s) \bX_i(s) \\
        (1-\pi_i(s)) \bX_i(s)\\
        \end{bmatrix}
\]
and \[
\pi_i(s) = \frac{e^{\bgamma^T\bX_i(s)}S(s)}
        {e^{\bbeta^T\bX_i(s)}F(s)+e^{\bgamma^T\bX_i(s)}S(s)}.
\]
Under model \eqref{GH1}, $\bU(t;\widehat{\btheta}_n,\widehat{\Lambda}_n)$
are expected to fluctuate randomly around 0.
Therefore along the line of \cite{Lin:93BKA},
we can construct an alternative goodness-of-fit test for the $j$th covariate based on the
test statistic
\[
K_j=\sup_{t\in[\delta,\tau-\delta]} \bU^T_j(t;\widehat{\bbeta}_n,\widehat{\Lambda}_n)
        \widehat{\text{Cov}}^{-1}\{\bU^T_j(t;\widehat{\bbeta}_n,\widehat{\Lambda}_n)\}
        \bU_j(t;\widehat{\bbeta}_n,\widehat{\Lambda}_n) , \ \ \ j=1,\cdots,p,
\]
where $\delta$ is a small positive number to avoid numerical problems at the edges,
and $\bU_j(\cdot)$ is the score process for the $j$th covariate. Similar to
\cite{Lin:93BKA}, the null distribution of the above test statistic can be
evaluated using a resampling approach and the p-value may be approximated
by the empirical proportions of the realizations of the null distribution exceeding
$K_j$.
The theoretical justification of this procedure, however, is challenging since
the partial likelihood function is not available under model \eqref{GH1}.
We are currently investigating this type of goodness-of-fit procedures
for general semiparametric survival models including model \eqref{GH1}.

To accommodate time-varying covariate effects on survival outcomes,
one can also extend the Cox model \eqref{Cox} through the use of time-varying
regression coefficients such that
\[
\lambda(t|\bX) = \lambda(t) e^{\bbeta^T(t)\bX},
\]
where $\bbeta(t)$ is a $p\times 1$ vector of unspecified functions of $t$.
Estimation and inference procedures for this
so-called varying-coefficient Cox model have been investigated by
several authors, including \cite{Zucker:90}, \cite{Murphy:91},
\cite{Murphy:93},
\cite{Martinussen:02SJS},
\cite{Winnett:03JRSSB}, \cite{Cai:03SJS}, \cite{Tian:05JASA},
and \cite{Peng:07BKA}, among others.
In general, nonparametric smoothing is required to estimate the time varying
coefficients. 
Note that for the case when $X$ is a one-dimensional binary covariate, as for the two arm  clinical trials, the time-varying regression coefficient model is completely nonparametric and specify any relationship between the two samples. For the general $k$-dimensional covariates, though,  
it may  be interesting to compare the performance of
the proposed method with that of the methods based on the varying-coefficient Cox model.

\vspace{1cm}
\begin{center}
ACKNOWLEDGMENTS
\end{center}
The authors are grateful to the COGA investigators
and Jean W. MacCluer for providing the
COGA data from GAW14, which was supported
in part by the NIH grant GM31575. The authors
thank Dr. William Rosenberger for making valuable
comments and suggestions, which lead to a considerable
improvement in the presentation of this manuscript.
The work of the
first author was supported
by the NIH grant R15CA150698.

\vspace{1cm}
\begin{center} 
APPENDIX
\end{center}

We introduce some notations that will be used throughout the appendix.
Let $\bO_i$ denote the observations for the $i$th subject consisting of 
$(Y_i,\Delta_i,\overline{\bX}_i)$. Let $\bP_n$ and $\bP$ be the empirical 
measure and the expectation of
$n$ i.i.d. observations $\bO_1,...,\bO_n$. That is,
for any measurable function $g(\bO)$,
\[
\bP_n[g(\bO)] = \frac{1}{n}\sum_{i=1}^n g(\bO_i),
\ \ \bP[g(\bO)]=E[g(\bO)].
\]

{\bf A.1. Proof of Lemma 1.} 
 Suppose that two sets of parameters,
$(\btheta, \Lambda)$ and $(\widetilde{\btheta},\widetilde{\Lambda})$, give
the same likelihood function for the observed data, i.e.,
\begin{equation}
\begin{split}
&       \biggl[\frac{e^{(\bbeta+\bgamma)^T\bX(Y)}\Lambda'(Y)}
        {e^{\bbeta^T\bX(Y)}F(Y)+e^{\bgamma^T\bX(Y)}S(Y)} \biggr]^\Delta e^{-\Lambda(Y|\overline{\bX}(Y)) } \\
&=       \biggl[\frac{e^{(\widetilde{\bbeta}+\widetilde{\bgamma})^T
        \bX(Y)}\widetilde{\Lambda}'(Y)}
        {e^{\widetilde{\bbeta}^T\bX(Y)}\widetilde{F}(Y)+e^{\widetilde{\bgamma}^T\bX(Y)}\widetilde{S}(Y)} \biggr]^\Delta e^{-\widetilde{\Lambda}(Y|\overline{\bX}(Y)) } \\
\end{split}
\label{ident}
\end{equation}
where $\widetilde{S}(t) = e^{-\widetilde{\Lambda}(t)}$, $\widetilde{F}(t)=1-
        \widetilde{S}(t)$, and
$\widetilde{\Lambda}(t|\overline{\bX}(t)) =
        \int_0^t \frac{e^{(\widetilde{\bbeta}+\widetilde{\bgamma})^T
        \bX(s)}}
        {e^{\widetilde{\bbeta}^T\bX(s)}\widetilde{F}(s)+e^{\widetilde{\bgamma}^T\bX(s)}\widetilde{S}(s)} d\widetilde{\Lambda}(s)$.
Let $\Delta=1$ and $Y=0$, we obtain
\[
(\bbeta-\widetilde{\bbeta})^T\bX(0) = \log \frac{\widetilde{\Lambda}'(0)}
        {\Lambda'(0)}.
\]
Then, condition (C1) gives $\bbeta=\widetilde{\bbeta}$ and
$\Lambda'(0) = \widetilde{\Lambda}'(0)$.
Because the equality \eqref{ident} holds for any $\overline{\bX}$, by letting
$\bX(s)=0, s\in[0,\tau]$ and $\Delta=0$, we obtain $\widetilde{\Lambda}(y)=\Lambda(y)$.
Finally, by choosing $\Delta=0$ and $Y=y$
and taking the logarithm and then the first derivative with respect
to $y$ in \eqref{ident}, we obtain
\[
\frac{e^{(\bbeta+\bgamma)^T\bX(y)}\Lambda'(y)}
        {e^{\bbeta^T\bX(y)}F(y)+e^{\bgamma^T\bX(y)}S(y)}
=\frac{e^{(\widetilde{\bbeta}+\widetilde{\bgamma})^T
        \bX(y)}\widetilde{\Lambda}'(y)}
        {e^{\widetilde{\bbeta}^T\bX(y)}\widetilde{F}(y)+e^{\widetilde{\bgamma}^T\bX(y)}\widetilde{S}(y)}.
\]
Again condition (C1) gives $\bgamma=\widetilde{\bgamma}$. The identifiability
of the parameters $(\btheta,\Lambda)$ is established.

{\bf A.2. Proof of Theorem 1.}
The proof of consistency consists of two major steps. In the first step, we prove that
$\widehat{\Lambda}_n(t)$ has an upper bound in $[0,\tau]$ with probability one. 
Therefore there exists a 
subsequence of $(\widehat{\btheta}_n,\widehat{\Lambda}_n)$ that converges to
$(\btheta^*,\Lambda^*)$. In the second step, we prove that $\btheta^*=\btheta_0$
and $\Lambda^*=\Lambda_0$.
 
 {\it Step 1.} We will prove the boundedness of $\widehat{\Lambda}_n(\tau)$
by contradiction. Recall that nonparametric log-likelihood takes the form
\[
l_n(\bbeta,\Lambda) = n\bP_n[R(\bO;\btheta,\Lambda)+\Delta \log\Lambda\{Y\}],
\]
where
\[
\begin{split}
R(\bO;\btheta,\Lambda) = & \Delta \left[(\bbeta+\bgamma)^T\bX(Y)
	- \log \left\{e^{\bbeta^T\bX(Y)}F(Y)
                + e^{\bgamma^T\bX(Y)}S(Y) \right\} \right] \\
	& - \int_0^Y \frac{e^{(\bbeta+\bgamma)^T\bX(y)}}
		{e^{\bbeta^T\bX(y)}F(y) + e^{\bgamma^T\bX(y)}S(y)}d\Lambda(y).\\
\end{split}
\]

Define $\widehat{\xi}_n=\widehat{\Lambda}_n(\tau)$ and $\widetilde{\Lambda}_n
(y) = \widehat{\Lambda}_n(y)/\widehat{\xi}_n$. It is obvious that $\widehat{\xi}_n$
maximizes the function
$l_n(\widehat{\btheta}_n,\xi \widetilde{\Lambda}_n)/n$. To prove
$\widehat{\Lambda}_n$ in $[0,\tau]$ is bounded, it is sufficient to
prove $\widehat{\xi}_n$ is bounded.
It is easy to see that
\[
\begin{split}
0 &\leq \frac{1}{n}l_n(\widehat{\btheta}_n,\widehat{\xi}_n\widetilde{\Lambda}_n)
        - \frac{1}{n}l_n(\widehat{\btheta}_n,\widetilde{\Lambda}_n) \\
 & =\bP_n \biggl[ \Delta \log \widehat{\xi}_n
       -\Delta \log \frac{e^{\widehat{\bbeta}_n^T\bX(Y)}\widehat{F}_n(Y)
        + e^{\widehat{\bgamma}_n^T\bX(Y)} \widehat{S}_n(Y)}{e^{\widehat{\bbeta}_n^T\bX(Y)}\widetilde{F}_n(Y)
        + e^{\widehat{\bgamma}_n^T\bX(Y)} \widetilde{S}_n(Y)}\\
 \ \ & -\int_0^{Y} e^{(\widehat{\bbeta}_n+
        \widehat{\bgamma}_n)^T\bX(t)}
        \biggl\{\frac{\widehat{\xi}_n}{e^{\widehat{\bbeta}_n^T\bX(t)}\widehat{F}_n(t)
        + e^{\widehat{\bgamma}_n^T\bX(t)} \widehat{S}_n(t)} \\
 \ \ & - \frac{1}{e^{\widehat{\bbeta}_n^T\bX(t)}\widetilde{F}_n(t)
        + e^{\widehat{\bgamma}_n^T\bX(t)} \widetilde{S}_n(t)} \biggr\}
        d\widetilde{\Lambda}_n(t)\biggr],\\
\end{split}
\]
where $(\widehat{F}_n,\widehat{S}_n)$ and $(\widetilde{F}_n,\widetilde{S}_n)$
are the distribution function and survival function corresponding to
$\widehat{\Lambda}_n$ and $\widetilde{\Lambda}_n$, respectively.

By conditions (C1) and (C4), we can show that
\[
\bP_n\biggl[ -\Delta \log \frac{e^{\widehat{\bbeta}_n^T\bX(Y)}\widehat{F}_n(Y)
        + e^{\widehat{\bgamma}_n^T\bX(Y)} \widehat{S}_n(Y)}{e^{\widehat{\bbeta}_n^T\bX(Y)}\widetilde{F}_n(Y)
        + e^{\widehat{\bgamma}_n^T\bX(Y)} \widetilde{S}_n(Y)}
        \biggr] \leq g_1,
\]
where $g_1$ is a constant.
Suppose that $\widehat{\xi}_n \rightarrow \infty$. According to conditions (C1) and (C4), we have
\[
 \bP_n  \biggl[ -\int_0^{Y} e^{(\widehat{\bbeta}_n+
        \widehat{\bgamma}_n)^T\bX(t)}
        \biggl\{\frac{\widehat{\xi}_n}{e^{\widehat{\bbeta}_n^T\bX(t)}\widehat{F}_n(t)
        + e^{\widehat{\bgamma}_n^T\bX(t)} \widehat{S}_n(t)} 
\leq -g_2 \widehat{\xi}_n+g_3
\]
for some positive constants $g_2$ and $g_3$.

It follows that
\[
0 
\leq \frac{1}{n}l_n(\widehat{\btheta}_n,\widehat{\xi}_n\widetilde{\Lambda}_n)
        - \frac{1}{n}l_n(\widehat{\btheta}_n,\widetilde{\Lambda}_n) 
\leq \log \widehat{\xi}_n -g_2\widehat{\xi}_n + g_3 
\rightarrow -\infty
\]
as $\widehat{\xi}_n\rightarrow \infty$. This contradicts to the definition of
$(\widehat{\btheta}_n,\widehat{\Lambda}_n)$. Note that the
above argument hold for every sample in the probability space
except a  set with zero probability.
Therefore we have shown that, with probability one, $\widehat{\Lambda}_n(\tau)$
is bounded for any sample size $n$.

Thus, by Helly's selection theorem, we can choose a further subsequence,  still 
indexed by $\{n\}$,  such that $\widehat{\btheta}_n\rightarrow\btheta^*$ and
$\widehat{\Lambda}_n$ weakly converges to $\Lambda^*$ with probability one.

{\it Step 2.} In this step, we will show that $\btheta^*=\btheta_0$ and $\Lambda^*=\Lambda_0$.
By differentiating $l_n(\btheta,\Lambda)$ with respect to $\Lambda\{Y_i\}$ and setting
it be zero, we can see that $\widehat{\Lambda}_n\{Y_i\}$ satisfies the
following equation.
\begin{equation}
\widehat{\Lambda}_n\{Y_i\} = \frac{\Delta_i}{n\bP_n[I(Y\geq y)
        Q(y,\bO;\widehat{\btheta}_n,\widehat{\Lambda}_n)
        ]}
        \bigg\vert_{y=Y_i},
\label{d1lambda}
\end{equation}
where
\[
\begin{split}
Q(y,\bO;\btheta,\Lambda)
= & \frac{\Delta S(Y) \{e^{\bbeta^T\bX(Y)}
                - e^{\bgamma^T\bX(Y)} \}}
	{e^{\bbeta^T\bX(Y)}F(Y)
                + e^{\bgamma^T\bX(Y)}S(Y)}
	+ \frac{e^{(\bbeta+\bgamma)^T
                \bX(y)}}{e^{\bbeta^T\bX(y)}F(y)
                + e^{\bgamma^T\bX(y)}S(y)}\\
           & -\int_y^{Y} \frac{e^{(\bbeta+\bgamma)^T\bX(s)}S(s)
        \{e^{\bbeta^T\bX(s)}
                - e^{\bgamma^T\bX(s)}\}}{\left\{e^{\bbeta^T\bX(s)}F(s)
                + e^{\bgamma^T\bX(s)}S(s)\right\}^2} d\Lambda(s).\\
 \end{split}
\]

In view of \eqref{d1lambda}, we construct another step function $\overline{\Lambda}_n(t)$ with
jumps only at the observed $Y_i$ and the jump size satisfies that
\[
\overline{\Lambda}_n\{Y_i\} = \frac{\Delta_i}{n\bP_n[I(Y\geq y)
        Q(y,\bO; \btheta_0,\Lambda_0) ]}
        \bigg\vert_{y=Y_i}.
\]
We verify that $\overline{\Lambda}_n(t)$ converges to
$\Lambda_0$ uniformly in $t\in[0,\tau]$ with probability one.
In Appendix A.4, we prove that the class
\[
\mathcal{F}_1=\{I(Y\geq y)Q(y,\bO; \btheta,\Lambda):y\in[0,\tau],\btheta\in\mathcal{B}_0,\Lambda\in\mathcal{A},\Lambda(0)=0\}
\]
is a bounded and P-Donsker class, where $\mathcal{A}=\{g: g \text{ is a nondecreasing function in }[0,\tau], g(\tau)\leq B_0\}$ and $B_0$ is a positive constant such that $\widehat{\Lambda}_n(\tau)
\leq B_0$ with probability one.
Since a P-Donsker class is also a Glivenko-Cantelli class, by the Glivenko-Cantelli
theorem in \cite{wellner:96}, $\overline{\Lambda}_n(t)$ uniformly converges 
to 
$
E[{I(Y\leq t)\Delta}/{\mu(Y)}
],
$
where
$
\mu(y) = E[I(Y\geq y)Q(y,\bO;\btheta_0,\Lambda_0)].
$

Denoting by $S_C(\cdot|\overline{\bX})$ the survival function of the 
censoring time $C$ given 
$\overline{\bX}$, we have
\[
\mu(y) 
= E\biggl[\frac{e^{(\bbeta_{0}+\bgamma_{0})^T\bX(y)-
	\Lambda_0(y|\overline{\bX}(y))}S_C(y|\overline{\bX}(y))}
	{e^{\bbeta_{0}^T\bX(y)}F_0(y)+e^{\bgamma_{0}^T\bX(y)}S_0(y)}\biggr],
\]
where $\Lambda_0(\cdot|\overline{\bX})$ is the true cumulative hazard function
of $T$ given $\overline{\bX}$, $F_0$ is the true baseline distribution 
function and $S_0$ is the true baseline survival function. 
Therefore,
\[
\begin{split}
E\biggl[\frac{I(Y\leq t)\Delta}{\mu(Y)}\biggr]
	& = E\biggl[\int_0^t\frac{e^{(\bbeta_{0}+\bgamma_{0})^T\bX(y)-
        \Lambda_0(y|\overline{\bX}(y))}S_C(y|\overline{\bX}(y))}
        {\mu(y)\{e^{\bbeta_{0}^T\bX(y)}F_0(y)+e^{\bgamma_{0}^T\bX(y)}S_0(y)\}}d\Lambda_0(y) \biggr]\\
	& = \int_0^t d\Lambda_0(y)=\Lambda_0(t).\\
\end{split}
\]
Consequently, we
conclude that $\overline{\Lambda}_n$ uniformly converges to $\Lambda_0$ in
$[0,\tau]$ with probability one.

By the construction of $\widehat{\Lambda}_n(t)$ and
$\overline{\Lambda}_n(t)$, we can see that $\widehat{\Lambda}_n(t)$ is absolutely 
continuous with respect to $\overline{\Lambda}_n(t)$ and 
\begin{equation}
\widehat{\Lambda}_n(t) = \int_0^t \frac{\bP_n[I(Y\geq y) Q(y,\bO;\btheta_0,\Lambda_0)]}
	{\bP_n[I(Y\geq y)Q(y,\bO;\widehat{\btheta}_n,\widehat{\Lambda}_n)]} d\overline{\Lambda}_n(y).
\label{LambdaStar}
\end{equation}
By taking limits on both sides of \eqref{LambdaStar}, we obtain that
\[
\Lambda^*(t)=\int_0^t \frac{\bP[I(Y\geq y) Q(y,\bO;\btheta_0,\Lambda_0)]}
	{\bP[I(Y\geq y)Q(y,\bO;\btheta^*,\Lambda^*)]} d\Lambda_0(y).
\]
Therefore, $\Lambda^*(t)$ is differentiable with respect to $\Lambda_0(t)$
so that $\Lambda^*(t)$ is differentiable with respect to $t$.
It follows that $d\widehat{\Lambda}_n(t)/d\overline{\Lambda}_n(t)$
converges to $d\Lambda^*(t)/d\Lambda_0(t)$ uniformly in $t\in[0,\tau]$. 
 
Note that
\begin{equation}
\begin{split}
n^{-1}& l_n(\widehat{\btheta}_n,\widehat{\Lambda}_n) - 
	n^{-1}l_n(\btheta_0,\overline{\Lambda}_n)\\
	& = \bP_n\biggl[\Delta\log \frac{\widehat{\Lambda}_n\{Y\}}
		{\overline{\Lambda}_n\{Y\}} \biggr]
	+ \bP_n [R(\bO;\widehat{\btheta}_n,\widehat{\Lambda}_n)
		- R(\bO; \btheta_0,\overline{\Lambda}_n)]\\
	& \geq 0.\\
\end{split}
\label{kullback}
\end{equation}
Since $\mathcal{B}_0\times \mathcal{A}$ is a Donsker class and the 
functionals $R(\bO;\btheta,\Lambda)$ are bounded Lipschitz functionals with respect to $\mathcal{B}_0\times \mathcal{A}$,
by the same arguments as in the proof of Donsker class for $\mathcal{F}_1$, the following class 
 \[
 \mathcal{F}_2=\{R(\bO; \btheta,\Lambda):\btheta\in\mathcal{B}_0,\Lambda\in\mathcal{A},
	\Lambda(0)=0,\Lambda(\tau)\leq B_0\}
\]
is P-Donsker and hence a Glivenko-Cantelli class. Therefore by letting $n\rightarrow \infty$ in 
\eqref{kullback}, we have
 \[
 0 \leq \bP\biggl[\log \biggl\{
 	\frac{\lambda^*(Y)^\Delta e^{R(\bO;\btheta^*,\Lambda^*)}}
		{\lambda_0(Y)^\Delta e^{R(\bO;\btheta_0,\Lambda_0)}} \biggr\} \biggr],
 \]
 which is the negative Kullback-Leibler information. Then it follows that, with
 probability one,
 \[
 \lambda^*(Y)^\Delta e^{R(\bO;\btheta^*,\Lambda^*)}
 =\lambda_0(Y)^\Delta e^{R(\bO;\btheta_0,\Lambda_0)}.
 \] 
 Therefore, from the identifiability
 result proved earlier, we obtain $\btheta^*=\btheta_0$ and $\Lambda^*=\Lambda_0$. 
 This completes the proof of Theorem 1. 
 
 {\bf A.3. Proof of Theorem 2.} 
We prove Theorem 2 by verifying the four conditions in Theorem 3.3.1
of \cite{wellner:96}. For this purpose, we first define a 
neighborhood of the true parameters $(\btheta_0,\Lambda_0)$,
denoted by 
\[
\mathcal{U} = \{(\btheta,\Lambda):||\btheta-\btheta_0||
+\sup_{t\in[0,\tau]}|\Lambda(t)-\Lambda_0(t)| < \epsilon_0\},
\]
for a very small constant $\epsilon_0$. 
Based on the consistency theorem, $(\widehat{\btheta}_n,
\widehat{\Lambda}_n)$ belongs to $\mathcal{U}$ with probability
close to 1 when the sample size $n$ is large enough.

For any one-dimensional submodel
 given as $\{\bbeta+\epsilon \bh_1,\bgamma+\epsilon\bh_2,
  \Lambda+\epsilon \int h_3 d\Lambda\}, (\btheta,\Lambda)\in\mathcal{U},\bH\equiv(\bh_1,\bh_2,h_3)\in\mathcal{H}$, we can derive the 
score function for a single observation $\bO$
\begin{equation}
\begin{split}
W(\bO;\btheta,\Lambda)[\bH] = & \Delta  \biggl[(\bh_1+\bh_2)^T\bX(Y)+h_3(Y) - \frac{R_2(Y,\bO;\btheta,\Lambda)[\bH]}{R_1(Y,\bO;\btheta,\Lambda)} \biggr]\\
&
-\int_0^Y \biggl[\frac{e^{(\bbeta+\bgamma)^T\bX(Y)}
	\{(\bh_1+\bh_2)^T\bX(Y) + h_3\}}{R_1(y,\bO;\btheta,\Lambda)} \\
	&-\frac{e^{(\bbeta+\bgamma)^T\bX(Y)}R_2(y,\bO;\btheta,\Lambda)[\bH]}
		{R_1^2(y,\bO;\btheta,\Lambda)}\biggr] d\Lambda,\\
\end{split} 
\label{scoreeq}
\end{equation}
where $R_1(y,\bO;\btheta,\Lambda)=e^{\bbeta^T\bX(y)}F(y) + e^{\bgamma^T\bX(y)}S(y)$
and
\[
\begin{split}
R_2(y,\bO; \btheta,\Lambda)[\bH] =&  e^{\bbeta^T\bX(y)}\left(F(y)\bh_1^T\bX(y)
	+  S(y) \int_0^y h_3 d\Lambda \right)\\ 
	&	+ e^{\bgamma^T\bX(y)}S(y)\left(\bh_2^T\bX(y) - \int_0^y h_3 d\Lambda \right).\\
\end{split}
\]
We define 
\[
U_n(\btheta,\Lambda)[\bH] = \bP_n\{W(\bO;\btheta,\Lambda)
	[\bH] \}
\]
and
\[
U(\btheta,\Lambda)[\bH] = \bP\{W(\bO;\btheta,\Lambda)
	[\bH] \}.
\]
Thus, it is easy to see that $U_n(\btheta,\Lambda)[\bH]$
and $U(\btheta,\Lambda)[\bH]$ are both maps from
$\mathcal{U}$ to $l^\infty(\mathcal{H})$ and
$\sqrt{n}\{U_n(\btheta,\Lambda)-U(\btheta,\Lambda)\}$ is an
empirical process in the space $l^\infty(\mathcal{H})$.
It is easy to see that $U_n(\widehat{\btheta}_n,\widehat{\Lambda}_n)
=0$ and $U(\btheta_0,\Lambda_0)=0$. 

We shall prove the theorem by verifying the following four
properties stated in Theorem 3.3.1 of \cite{wellner:96}.
\begin{enumerate}
 \item[(P1)] $\sqrt{n}(U_n-U)(\widehat{\btheta}_n,\widehat{\Lambda}_n)
	- \sqrt{n}(U_n-U)(\btheta_0,\Lambda_0)
	= o_P(1+\sqrt{n}||\widehat{\btheta}_n-\btheta_0||
	+ \sqrt{n}\sup_{y\in[0,\tau]}|\widehat{\Lambda}_n(y)
	-\Lambda_0(y)|)$.
\item[(P2)] $\sqrt{n}(U_n-U)(\btheta_0,\Lambda_0)$ converges
to a tight random element $\bxi$.
\item[(P3)] $U(\btheta,\Lambda)$ is Frechet-differentiable  at 
$(\btheta_0,\Lambda_0)$.
\item[(P4)] The derivative of $U(\btheta,\Lambda)$ at
$(\btheta_0,\Lambda_0)$, denoted by $U'(\btheta_0,\Lambda_0)$
is continuously invertible.
\end{enumerate}

To prove property (P1), we make use of Lemma 3.3.5 of 
\cite{wellner:96}. Based on the explicit expression in \eqref{scoreeq},
$W(\bO;\btheta,\Lambda)[\bH]$
is continuously differentiable with respect to $\btheta$ and
\[
\bigg\lVert \frac{dW(\bO;\btheta,\Lambda)}{d\btheta}\bigg\rVert 
\leq g_4, 
\]
where $g_4$ is a positive constant. Furthermore, 
\[
|W(\bO;\btheta,\Lambda_1)[\bH]-W(\bO;\btheta,\Lambda_2)[\bH]|
\leq g_5 \biggl\{|\Lambda_1(Y)-\Lambda_2(Y)|
	+ \int_0^\tau |\Lambda_1(y)-\Lambda_2(y)|dy \biggr\}
\]
for some positive constant $g_5$.
Therefore,
\[
\sup_{\bH\in\mathcal{H}} E\biggl[\left\{W(\bO;\btheta,\Lambda)[\bH]
	-W(\bO;\btheta_0,\Lambda_0)[\bH]\right\}^2\biggr]
\]
converges to zero if $||\btheta-\btheta_0||+\sup_{y\in[0,\tau]}
|\Lambda(y)-\Lambda_0(y)|\rightarrow 0$. 
In addition, by the same arguments as in the proof of Donsker 
class for $\mathcal{F}_1$, the class
\[
\mathcal{F}_3=\{W(\bO;\btheta,\Lambda)[\bH]
-W(\bO;\btheta_0,\Lambda_0)[\bH]:
	(\btheta,\Lambda)\in\mathcal{U}, \bH\in
	\mathcal{H}\}
\]
is P-Donsker. Therefore, according to Lemma 3.3.5 of \cite{wellner:96}, property (P1) holds.

Property (P2) holds again because of the P-Donsker property of the
class
\[
\{W(\bO;\btheta_0,\Lambda_0)[\bH]:\bH
\in\mathcal{H}\}.
\]
Furthermore, the limit random elements $\bxi$ is a Gaussian process
indexed by $\bH\in\mathcal{H}$ and the
covariance between $\bxi(\bH_1)$ and
$\bxi(\bH_2)$ is equal to
\[
E\biggl[W(\bO;\btheta_0,\Lambda_0)[\bH_1] \times 
	W(\bO;\btheta_0,\Lambda_0)[\bH_2] \biggr].
\]

The Frechet differentiability in (P3) can be directly verified by using the smoothness of $U(\btheta,\Lambda)$. The derivative 
of $U(\btheta,\Lambda)$ at $(\btheta_0,\Lambda_0)$,
denoted by $U'(\btheta_0,\Lambda_0)$ is a map from the space
\[
\{(\btheta-\btheta_0,\Lambda-\Lambda_0): (\btheta,\Lambda)\in\mathcal{U}\}
\]
to $l^\infty(\mathcal{H})$.

It remains to show that $U'$ is continuously
invertible at $(\btheta_0,\Lambda_0)$. Follow the argument 
in the Appendix of \cite{ZengLin:07JRSSB}, it suffices to prove 
that for any one-dimensional submodel
 given as $\{\bbeta_{0}+\epsilon \bh_1,\bgamma_0+\epsilon\bh_2,
  \Lambda_0+\epsilon \int h_3 d\Lambda_0\}, \bH
\in\mathcal{H}$, the Fisher information along this submodel is nonsingular. If the Fisher information along this submodel is
singular, the score function along this submodel is
zero with probability one. We will show that 
$W(\bO;\btheta_0,\Lambda_0)[\bH]=0$ yields 
that $\bh_1={\bf 0},\bh_2={\bf 0}$, and $h_3=0$. 
We follow the ideas of proving the identifiability 
in the proof of Theorem 1. 
Let $\Delta=1$ and $Y=0$, we obtain 
$\bh_1^T\bX(0)+h_3(0)=0$. Conditions (C1) gives $\bh_1=0$ and $h_3(0)=0$. 
Let $\Delta=0$ and $\bX(s)=0, s\in[0,\tau]$, we obtain $\int_0^t h_3 d\Lambda_0=0$ for 
any $t\in[0,\tau]$. Similarly, let $\Delta=1$ and $\bX(s)=0, s\in[0,\tau]$,
we obtain $h_3(t)+\int_0^t h_3d\Lambda_0=0$. Therefore, $h_3(t)=0$ for any $t\in[0,\tau]$. 
Let $\Delta=0$ and $Y=y$ and then take the first derivative with respect to $y$ in $W(\bO;\btheta_0,\Lambda_0)[\bH]$,
we obtain
\[
\bh_2^T\bX(y)\frac{e^{\bbeta_{0}^T\bX(y)}F_0(y)}{R_1(y,\bO;\btheta_0,\Lambda_0)}=0
\]
for any $y\in [0,\tau]$. Immediately, we have $\bh_2=0$. 
We have thus proved nonsingularity of the Fisher information matrix along 
any nontrivial submodel. Hence, property (P4) holds.

We now have verified properties (P1)-(P4), Theorem 3.3.1 of
\cite{wellner:96} concludes that $\sqrt{n}(\widehat{\btheta}_n
-\bbeta_0,\widehat{\Lambda}_n-\Lambda_0)$ weakly converges
to a tight Gaussian random element $-U'^{-1}\bxi$ in $l^\infty(\mathcal{H})$. Moreover, it can be shown that 
$\widehat{\btheta}_n$ is an asymptotic linear estimator for
$\btheta_0$ and that the corresponding influence functions 
are on the space spanned by the score functions.
This implies that $\widehat{\btheta}_n$ is semiparametrically
efficient by the semiparametric efficiency theory \citep{Bickel:93}.

{\bf A.4. Donsker Property of $\mathcal{F}_1$.}
In this appendix, we prove that the following  class
\[
\mathcal{F}_1=\{I(Y\geq y)Q(y,\bO; \btheta,\Lambda):y\in[0,\tau],\btheta\in\mathcal{B}_0,\Lambda\in\mathcal{A},\Lambda(0)=0\},
\]
is P-Donsker.
To show that $\mathcal{F}_1$ is P-Donsker, we first prove that the class
\[
\mathcal{F}=\{Q(y,\bO; \btheta,\Lambda):y\in[0,\tau],\btheta\in\mathcal{B}_0,\Lambda\in\mathcal{A},\Lambda(0)=0,\Lambda(\tau)\leq B_0\}
\]
is P-Donsker.
Using condition (C2), it is easy to show that $Q(y,\bO;\btheta,\Lambda)$ is 
bounded and continuously
differentiable with respect to $\btheta$
for any $\btheta\in\mathcal{B}_0$
and
\[
\bigg\lVert \frac{dQ(y,\bO;\btheta,\Lambda)}{d\btheta}\bigg\rVert 
\leq g_6, 
\]
where $g_6$ is a positive constant. In addition, for any $\Lambda_1$ and $\Lambda_2\in
\mathcal{A}$ there exist a positive constant $g_7$ such that
\[
\begin{split}
|Q & (y,\bO;\btheta,\Lambda_1)-Q(y,\bO;\btheta,\Lambda_2)| \\
	&\leq g_7 \left\{|\Lambda_1(Y)-\Lambda_2(Y)|
		+ |\Lambda_1(y)-\Lambda_2(y)|
		+ \int_0^\tau |\Lambda_1(t)-\Lambda(t)|dt \right\}.\\
\end{split}
\]
Therefore, by the mean-value theorem, we can show that for any
$(y,\btheta,\Lambda)$ and $(y,\widetilde{\btheta},\widetilde{\Lambda})$
in $[0,\tau]\times \mathcal{B}_0\times\mathcal{A}$,
\[
\begin{split}
|Q & (y,\bO;\btheta,\Lambda)-Q(y,\bO;\widetilde{\btheta},\widetilde{\Lambda})| \\
	& \leq g_8 \biggl\{||\bbeta-\widetilde{\bbeta}||  + |\Lambda_1(Y)-\Lambda_2(Y)| \\
	& 	+ |\Lambda_1(y)-\Lambda_2(y)|
		+ \int_0^\tau |\Lambda_1(t)-\Lambda(t)|dt \biggr\}\\
\end{split}
\]
holds for a positive constant $g_8$. 
Since $[0,\tau]\times\mathcal{B}_0\times
\mathcal{A}$ and  
$\{H(y): y\in[0,\tau], H\in \mathcal{A}, H(0)=0, H(\tau)\leq B_0\}$ are both Donsker
classes, we conclude that $\mathcal{F}$ is P-Donsker
according to Theorems 2.7.5 and 2.5.6 in \cite{wellner:96}
and the preservation of the Donsker property under the product and the summation.
Similarly, since $\{I(Y\geq y): y\in[0,\tau]\}$
is P-Donsker, $\mathcal{F}_1$ is also P-Donsker.

\label{s:literature}
\bibliographystyle{asa}
\bibliography{cite_Diao}

\begin{thebibliography}{27}
\newcommand{\enquote}[1]{``#1''}
\expandafter\ifx\csname natexlab\endcsname\relax\def\natexlab#1{#1}\fi

\bibitem[{Bennett({1983})}]{Bennett:83}
Bennett, S. ({1983}), \enquote{{Analysis of survival data by the proportional
  odds model},} \textit{{Statistics in Medicine}}, {2}, {273--277}.

\bibitem[{Bickel et~al.({1993, Ch. 3})Bickel, Klaassen, Ritov, and
  Wellner}]{Bickel:93}
Bickel, P.~J., Klaassen, C. A.~J., Ritov, Y., and Wellner, J.~A. ({1993, Ch.
  3}), \textit{Efficient and Adaptive Estimation for Semiparametric Models},
  Baltimore: Johns Hopkins University Press.

\bibitem[{Broyden(1970)}]{Broyden:1970}
Broyden, C.~G. (1970), \enquote{The convergence of a class of double rank
  minimization algorithms: 2: The new algorithm,} \textit{IMA Journal of
  Applied Mathematics}, 6, 222--231.

\bibitem[{Cai and Sun(2003)}]{Cai:03SJS}
Cai, Z. and Sun, Y. (2003), \enquote{Local linear estimation for time-dependent
  coefficients in CoxÕs regression models,} \textit{Scandinavian Journal of
  Statistics}, 30, 93--111.

\bibitem[{Cox({1972})}]{Cox:72}
Cox, D.~R. ({1972}), \enquote{{Regression model and life-tables (with
  Discussion)},} \textit{{Journal of the Royal Statistical Society, Series B}},
  {34}, {187--220}.

\bibitem[{Diao and Lin(2010)}]{Diao:09GE}
Diao, G. and Lin, D.~Y. (2010), \enquote{{Variance-componens methods for
  linkage and association analysis of ordinal traits in general pedigrees},}
  \textit{Genetic Epidemiology}, {34}, 232--237.

\bibitem[{Fletcher(1970)}]{Fletcher:1970}
Fletcher, R. (1970), \enquote{A new approach to variable metric algorithms,}
  \textit{The Computer Journal}, 13, 317--322.

\bibitem[{{Gastrointestinal Tumor Study Group}(1982)}]{GTSG:82}
{Gastrointestinal Tumor Study Group} (1982), \enquote{A comparison of
  combination chemotherapy and combined modality therapy for locally advanced
  gastric carcinoma,} \textit{Cancer}, 49, 1771--1777.

\bibitem[{Goldfarb(1970)}]{Goldfarb:1970}
Goldfarb, D. (1970), \enquote{A family of variable metric methods derived by
  variational means,} \textit{Mathematics of Computation}, 24, 23--26.

\bibitem[{Hasin(2003)}]{Hasin:03}
Hasin, D. (2003), \enquote{Classification of alcohol use disorders,}
  \textit{Alcohol Research \& Health}, 27, 5--17.

\bibitem[{Lin et~al.({1993})Lin, Wei, and Ying}]{Lin:93BKA}
Lin, D.~Y., Wei, L.~J., and Ying, Z. ({1993}), \enquote{{Checking the Cox model
  with cumulative sums of martingale-based residuals},} \textit{{Biometrika}},
  {80}, {557--572}.

\bibitem[{Martinussen et~al.(2002)Martinussen, Scheike, and
  Skovgaard}]{Martinussen:02SJS}
Martinussen, T., Scheike, T.~H., and Skovgaard, I.~M. (2002),
  \enquote{Efficient estimation of fixed and time-varying covariate effects in
  multiplicative intensity models,} \textit{Scandinavian Journal of
  Statistics}, 29, 57--74.

\bibitem[{Murphy(1993)}]{Murphy:93}
Murphy, S.~A. (1993), \enquote{Testing for a time dependent coefficient in
  Cox¿s regression model,} \textit{Scandinavian Journal of Statistics}, 20,
  35--50.

\bibitem[{Murphy et~al.(1997)Murphy, Rossini, and van~der
  Vaart}]{Murphy:97JASA}
Murphy, S.~A., Rossini, A.~J., and van~der Vaart, A.~W. (1997),
  \enquote{Maximal likelihood estimation in the proportional odds model,}
  \textit{Journal of the American Statistical Association}, 92, 968--976.

\bibitem[{Murphy and Sen(1991)}]{Murphy:91}
Murphy, S.~A. and Sen, P.~K. (1991), \enquote{Time-dependent coefficients in a
  Cox-type regression model,} \textit{Stochastic Processes and their
  Applications}, 39, 153--180.

\bibitem[{Murphy and van~der Vaart({2000})}]{Murphy:00}
Murphy, S.~A. and van~der Vaart, A.~W. ({2000}), \enquote{On the profile
  likelihood,} \textit{{ Journal of the American Statistical Association}}, 95,
  {449--465}.

\bibitem[{Palmer et~al.(1999)Palmer, Katrina, and Burton}]{Palmer:99}
Palmer, L.~J., Katrina, J.~T., and Burton, P.~R. (1999), \enquote{Genome-wide
  linkage analysis using genetic variance components of alcohol
  dependency-associated censored and continuous traits,} \textit{Genetic
  Epidemiology}, 17(Suppl. 1), S283--S288.

\bibitem[{Peng and Huang(2007)}]{Peng:07BKA}
Peng, L. and Huang, Y. (2007), \enquote{Survival analysis with temporal
  covariate effects,} \textit{Biometrika}, 94, 719--733.

\bibitem[{Press et~al.(1992)Press, Teukolsky, Vetterling, and
  Flannery}]{Press:92}
Press, W.~H., Teukolsky, S.~A., Vetterling, W.~T., and Flannery, B.~P. (1992),
  \textit{Numerical Recipes in C: The Art of Scientific Computing, Second
  Edition}, Cambridge: Cambridge University Press.

\bibitem[{Shanno(1970)}]{Shanno:1970}
Shanno, D.~F. (1970), \enquote{Conditioning of quasi-Newton methods for
  function minimization,} \textit{Mathematics of Computation}, 24, 647--650.

\bibitem[{Tian et~al.(2005)Tian, Zucker, and Wei}]{Tian:05JASA}
Tian, L., Zucker, D., and Wei, L.~J. (2005), \enquote{On the Cox model with
  time-varying regression coefficients,} \textit{Journal of the American
  Statistical Association}, 100, 172--183.

\bibitem[{van~der Vaart and Wellner(1996)}]{wellner:96}
van~der Vaart, A. and Wellner, J. (1996), \textit{Weak Convergence and
  Empirical Processes: With Applications to Statistics}, New York:
  Springer-Verlag.

\bibitem[{Wang et~al.(2006)Wang, Ye, and Zhang}]{Wang:06GE}
Wang, X., Ye, Y., and Zhang, H. (2006), \enquote{Family-based association test
  for ordinal traits adjusting for covariates,} \textit{Genetic Epidemiology},
  30, 728--736.

\bibitem[{Winnett and Sasieni(2003)}]{Winnett:03JRSSB}
Winnett, A. and Sasieni, P. (2003), \enquote{Iterated residuals and
  time-varying covariate effect in Cox regression,} \textit{Journal of the
  Royal Statistical Society: Series B}, 65, 473--488.

\bibitem[{Yang and Prentice({2005})}]{Yang:05BKA}
Yang, S. and Prentice, R. ({2005}), \enquote{{Semiparametric analysis of
  short-term and long-term hazard ratios with two-sample survival data},}
  \textit{{Biometrika}}, {92}, {1--17}.

\bibitem[{Zeng and Lin(2007)}]{ZengLin:07JRSSB}
Zeng, D. and Lin, D.~Y. (2007), \enquote{Maximum likelihood estimation in
  semiparametric regression models with censored data (with discussion),}
  \textit{Journal of the Royal Statistical Society: Series B}, 69, 507--564.

\bibitem[{Zucker and Karr(1990)}]{Zucker:90}
Zucker, D.~M. and Karr, A.~F. (1990), \enquote{Nonparametric survival analysis
  with timedependent covariate effects: A penalized partial likelihood
  approach,} \textit{Annals of Statistics}, 18, 329--353.

\end{thebibliography}

\newpage
{\small
\begin{center}
Table 1. Summary statistics for the simulation studies based on 1,000 replications\\
\vspace{2mm}
\begin{tabular}{ccrrrrrrrrr}\hline
$n$     &Par    &Est    &SE     &SEE    &CP     &       &Est    &SE     &SEE&CP\\
\hline
        &       &\multicolumn{4}{c}{$(\beta,\gamma)=(-0.5,0.5)$} & &\multicolumn{4}{c}{$(\beta,\gamma)=(-0.5,0.0)$}\\
100     &$\beta$ &-0.511 &0.413 &0.412 &0.956 &&-0.506 &0.407 &0.409 &0.958\\
        &$\gamma$&0.465 &0.570 &0.556 &0.938 &&-0.04 &0.564 &0.568 &0.944\\
        &$\Lambda(0.5)$&0.508 &0.086 &0.085 &0.946 &&0.507 &0.085 &0.086 &0.962\\
        &$\Lambda(1.0)$&1.019 &0.145 &0.146 &0.954 &&1.019 &0.146 &0.149 &0.959\\
        &$\beta_{PH}$&-0.116 &0.217 &0.209 &- &&-0.317 &0.219 &0.211 &0.926\\
        &$\beta_{PO}$&-0.294 &0.327 &0.320 &- &&-0.511 &0.326 &0.322 &0.944\\
200     &$\beta$ &-0.512 &0.291 &0.288 &0.954 &&-0.507 &0.287 &0.286 &0.955\\
        &$\gamma$ &0.496 &0.400 &0.389 &0.940 &&-0.007 &0.401 &0.399 &0.950\\
        &$\Lambda(0.5)$&0.504 &0.059 &0.059 &0.954 &&0.504 &0.058 &0.060 &0.953\\
        &$\Lambda(1.0)$&1.012 &0.104 &0.101 &0.947 &&1.012 &0.104 &0.104 &0.953\\
        &$\beta_{PH}$&-0.107 &0.153 &0.147 &- &&-0.308 &0.154 &0.148 &-\\
        &$\beta_{PO}$&-0.287 &0.231 &0.225 &- &&-0.504 &0.231 &0.226 &0.948\\
        &       &\multicolumn{4}{c}{$(\beta,\gamma)=(0.0,0.5)$} & &\multicolumn{4}{c}{$(\beta,\gamma)=(0.5,0.5)$}\\
100     &$\beta$ &-0.012 &0.406 &0.405 &0.954 &&0.495 &0.414 &0.409 &0.945\\
        &$\gamma$ &0.490 &0.570 &0.563 &0.934 &&0.512 &0.587 &0.585 &0.947\\
        &$\Lambda(0.5)$&0.510 &0.087 &0.085 &0.952 &&0.509 &0.087 &0.087 &0.954\\
        &$\Lambda(1.0)$&1.023 &0.146 &0.148 &0.958 &&1.027 &0.147 &0.151 &0.959\\
        &$\beta_{PH}$&0.188 &0.211 &0.210 &- &&0.499 &0.216 &0.214 &0.952\\
        &$\beta_{PO}$&0.202 &0.321 &0.319 &- &&0.707 &0.327 &0.325 &-\\
200     &$\beta$ &-0.009 &0.284 &0.282 &0.956 &&0.496 &0.287 &0.285 &0.962\\
        &$\gamma$ &0.501 &0.398 &0.395 &0.947 &&0.503 &0.410 &0.411 &0.944\\
        &$\Lambda(0.5)$&0.506 &0.059 &0.060 &0.952 &&0.505 &0.059 &0.061 &0.957\\
        &$\Lambda(1.0)$&1.014 &0.104 &0.102 &0.944 &&1.015 &0.104 &0.105 &0.957\\
        &$\beta_{PH}$&0.193 &0.149 &0.147 &- &&0.498 &0.151 &0.150 &0.946\\
        &$\beta_{PO}$&0.207 &0.227 &0.225 &- &&0.706 &0.228 &0.229 &-\\
\hline
\end{tabular}
\end{center}
\noindent Par, the parameter to be estimated; Est, the average estimate; SE, the sample standard
deviation of the estimates; SEE, the average standard error; CP, the coverage probability of
the nominal 95\% confidence intervals.
}

\newpage
\begin{center}
Table 2. Empirical size/power of the Wald test at significance level of 0.05 based on 1,000 replications\\
\vspace{2mm}
\begin{tabular}{rrccccc}\hline
$\beta$ &$\gamma$ &H1 &H2 &H3 &H4 &H5\\
\hline 
0.0& 0.0& 0.040& 0.052& 0.050& 0.050& 0.059\\
-0.5& -0.5& 0.434& 0.246& 0.860& 0.052& 0.917\\
-0.5& -0.4& 0.439& 0.184& 0.801& 0.053& 0.874\\
-0.5& -0.3& 0.428& 0.142& 0.723& 0.059& 0.815\\
-0.5& -0.2& 0.429& 0.086& 0.638& 0.071& 0.741\\
-0.5& -0.1& 0.437& 0.051& 0.563& 0.096& 0.656\\
-0.5& 0.0& 0.438& 0.050& 0.499& 0.137& 0.544\\
-0.5& 0.1& 0.431& 0.062& 0.440& 0.166& 0.447\\
-0.5& 0.2& 0.437& 0.089& 0.396& 0.220& 0.345\\
-0.5& 0.3& 0.432& 0.137& 0.372& 0.268& 0.254\\
-0.5& 0.4& 0.428& 0.189& 0.363& 0.328& 0.179\\
-0.5& 0.5& 0.433& 0.262& 0.362& 0.398& 0.129\\
-0.4& 0.5& 0.304& 0.258& 0.287& 0.341& 0.074\\
-0.3& 0.5& 0.195& 0.253& 0.217& 0.264& 0.063\\
-0.2& 0.5& 0.104& 0.266& 0.191& 0.228& 0.075\\
-0.1& 0.5& 0.055& 0.266& 0.212& 0.176& 0.154\\
0.0& 0.5& 0.044& 0.263& 0.272& 0.139& 0.265\\
0.1& 0.5& 0.056& 0.261& 0.359& 0.109& 0.400\\
0.2& 0.5& 0.099& 0.245& 0.476& 0.086& 0.563\\
0.3& 0.5& 0.174& 0.248& 0.632& 0.065& 0.718\\
0.4& 0.5& 0.308& 0.231& 0.742& 0.057& 0.840\\
0.5& 0.5& 0.417& 0.223& 0.851& 0.047& 0.911\\
\hline
\end{tabular}
\end{center}
\clearpage

\newpage
\begin{center}
Table 3. Mean squared errors of the proposed NPMLEs and the pseudo maximum
likelihood estimators (PMLEs) of \cite{Yang:05BKA} for $(\beta,\gamma)$\\
\vspace{2mm}
\begin{tabular}{cccccccccc}\hline
        &       &\multicolumn{2}{c}{PMLE} & &\multicolumn{2}{c}{NPMLE} &
        &\multicolumn{2}{c}{PMLE/NPMLE}\\
\cline{3-4} \cline{6-7} \cline{9-10}
$n$     &$(\beta,{\gamma})$ &$\widehat{\beta}$ &$\widehat{\gamma}$
& &$\widehat\beta$ &$\widehat\gamma$& &$\widehat\beta$ &$\widehat\gamma$\\
\hline
100 &(-0.5,0.5) &0.090 &0.108 &&0.073 &0.111 &&1.242 &0.978\\
        &(-0.5,0.0) &0.085 &0.114 &&0.061 &0.105 &&1.390 &1.084\\
        &(0.0,0.5) &0.069 &0.107 &&0.063 &0.110 &&1.101 &0.967\\
        &(0.5,0.5) &0.088 &0.144 &&0.067 &0.133 &&1.314 &1.087\\
200 &(-0.5,0.5) &0.048 &0.060 & &0.036 &0.054 & &1.360 &1.107\\
        &(-0.5,0.0) &0.041 &0.061 & &0.031 &0.0543 & &1.310 &1.119\\
        &(0.0,0.5) &0.030 &0.050 & &0.030 &0.0516 & &1.025 &0.974\\
        &(0.5,0.5) &0.035 &0.064 & &0.030 &0.0598 & &1.152 &1.068\\
\hline
\end{tabular}
\end{center}

\newpage
\begin{figure}
\setlength{\abovecaptionskip}{-10pt}
\setlength{\belowcaptionskip}{10pt}
\begin{center}
\includegraphics[angle=0,width=6 in, totalheight=4 in]
{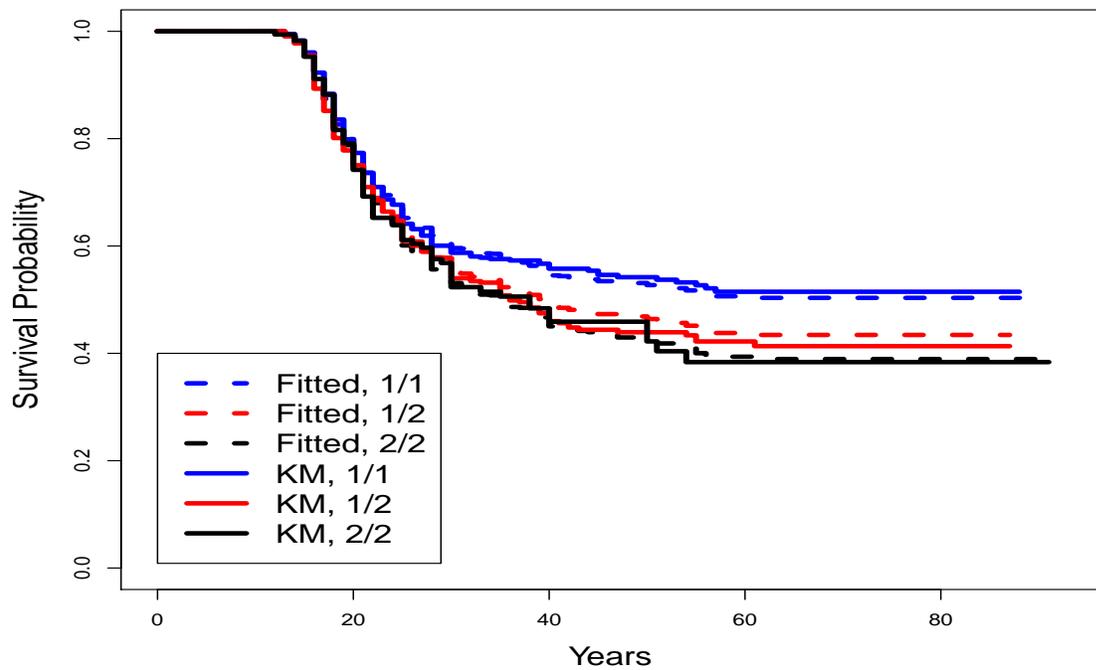}
\caption{Kaplan-Meier and model-fitted survival curves from the COGA study.}
\end{center}
\end{figure}

\begin{figure}
\setlength{\abovecaptionskip}{-10pt}
\setlength{\belowcaptionskip}{10pt}
\begin{center}
\includegraphics[angle=0,width=6 in, totalheight=4 in]
{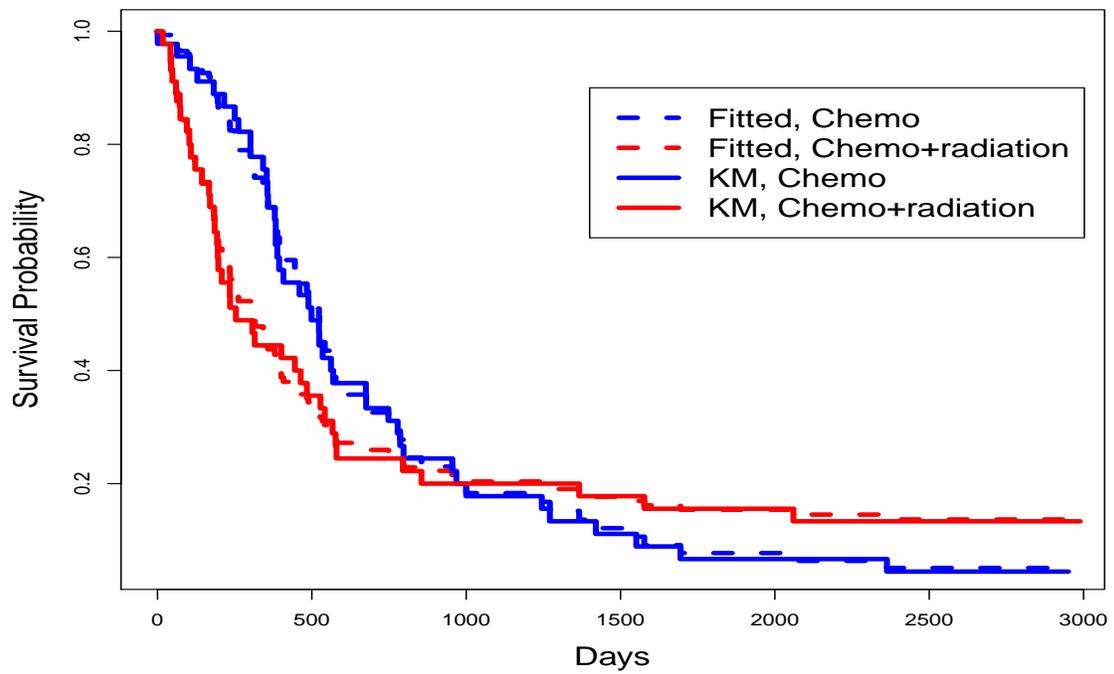}
\caption{Kaplan-Meier and model-fitted survival curves from the Gastrointestinal
tumor study.}
\end{center}
\end{figure}

\end{document}